\documentclass[10pt,aps,pra,twocolumn,groupedaddress,superscriptaddress,showpacs]{revtex4-1}
\usepackage{stmaryrd}
\usepackage{amssymb,lipsum}
\usepackage{amsmath,har2nat}
\usepackage{amsfonts,dsfont,amsthm,footnote}
\usepackage{color}
\usepackage{epsfig,graphicx}
\usepackage{bm,bibunits}
\usepackage{rotating}
\usepackage{mathrsfs}
\usepackage{hyperref}

\usepackage{amsmath,mathrsfs,amsbsy,color,graphicx,bm,amsthm,amsfonts,dsfont,chngcntr}
\usepackage[english]{babel}
\usepackage{amssymb,lipsum,attachfile}
\usepackage{amsmath,epsfig,epstopdf}
\usepackage{amsfonts,dsfont,amsthm}
\usepackage{hyperref}
\usepackage[normalem]{ulem}

\newcommand{\be}{\begin{equation}}
\newcommand{\ee}{\end{equation}}
\newcommand{\bea}{\begin{eqnarray}}
\newcommand{\eea}{\end{eqnarray}}
\newcommand{\ket}{\rangle}
\newcommand{\bra}{\langle}

\newcommand{\I}{\mathds{1}}
\newcommand{\ra}{\rightarrow}

\newcommand{\m}{\mathcal}
\def\C#1{\mathcal #1}
\def\B#1{\mathbb #1}

\newcommand{\T}[2]{\textsf{#1#2}}

\begin{document}
\newtheorem{theorem}{Theorem}
\newtheorem{proposition}[theorem]{Proposition}
\newtheorem{corollary}[theorem]{Corollary}
\newtheorem{open problem}[theorem]{Open Problem}
\newtheorem{definition}{Definition}
\newtheorem{remark}{Remark}
\newtheorem{example}{Example}

\title{Quasi-exact quantum computation}

\author{Dong-Sheng Wang}
\affiliation{Institute for Quantum Computing and Department of Physics and Astronomy, \\ University of Waterloo, Waterloo, ON, Canada}
\author{Guanyu Zhu}
\affiliation{IBM T. J. Watson Research Center, Yorktown Heights, NY, 10598, United States}
\author{Cihan Okay}
\affiliation{Stewart Blusson Quantum Matter Institute and Department of Physics and Astronomy, University of British Columbia, Vancouver, BC, Canada}
\author{Raymond Laflamme}
\affiliation{Institute for Quantum Computing and Department of Physics and Astronomy, \\ University of Waterloo, Waterloo, ON, Canada}

\begin{abstract}
  We study quasi-exact quantum error correcting codes
  and quantum computation with them.
  A quasi-exact code is an approximate code such that
  it contains a finite number of scaling parameters,
  the tuning of which can flow it to corresponding exact codes,
  serving as its fixed points.
  The computation with a quasi-exact code cannot realize any logical gate to arbitrary accuracy.
  To overcome this, the notion of quasi-exact universality is proposed,
  which makes quasi-exact quantum computation a feasible model especially for executing moderate-size algorithms.
  We find that the incompatibility between universality and transversality of the set of logical gates
  does not persist in the quasi-exact scenario.
  A class of covariant quasi-exact codes is defined which proves to support
  transversal and quasi-exact universal set of logical gates for $SU(d)$.
  This work opens the possibility of quantum computation with quasi-exact
  universality, transversality, and fault tolerance.
\end{abstract}

\maketitle


\section{Introduction}\label{sec:intr}

Fault-tolerant quantum computation~\cite{Shor96,AB97,KLZ98,Pre98}
requires quantum codes with error-correction features.
One way to introduce fault tolerance is to use transversal gates~\cite{KLZ96,EK09}.
Unfortunately, it is not possible to find a universal set of logical gates that are transversal~\cite{EK09,ZCC11,CCC+08}.
Such a `no-go' result is further explored for topological stabilizer codes
with local stabilizers~\cite{Got98,Kit03},
in which setting transversal gates are extended to
gates described by finite-depth local unitary circuits~\cite{BK13,PY15,BBK+16}.

In this work, we explore methods to circumvent the no-go theorems.
Methods in literature mostly focus on relaxing transversality.
A common method is to use fault-tolerant measurements,
such as magic-state injection~\cite{BK05}
and code switching~\cite{PR13}.
Another method is to use concatenation~\cite{KL96,JL14},
which achieves fault tolerance without fixed transversality.
Quantum computing by non-abelian anyons~\cite{NSS+08},
realized in lattice systems,
implement braidings by finite-depth local circuits~\cite{KKR10,BD12,ZHB17},
hence not transversal.

Transversal gates serve as a primary class of fault-tolerant gates
since they do not couple local subsystems,
on which independent error correction can be performed,
hence do not spread out local errors.
Instead of transversality, in this work
we relax the requirements on error correction and universality.
As Eastin-Knill originally suggested~\cite{EK09},
using approximate error correcting codes instead of exact codes
could be a viable approach to avoid the no-go constraint.
In this work, we explore this direction in depth.

Approximate codes~\cite{LNC+97,CGS05,RW05,BO10,BO11,Ben11,NM10,MN12,Pre00,HNP+17,FCS+18,FNA+19,WA19}
have been studied in various contexts,
and in particular, conditions for approximate error correction have been studied~\cite{BO10,BO11,Ben11,NM10,MN12}.
However, the leftover error after error correction, i.e., recovery error, may not be small.
To achieve accurate quantum computation,
we have to restrict the setting of approximate codes as follows.
We call an approximate code to be `quasi-exact' if
there exist a finite number of tunable \emph{scaling parameters} such that
it can approach exact codes as close as possible.
This also implies that the recovery error,
which is the accuracy parameter of a code,
can be arbitrarily small with well-defined scaling behaviors.
Examples of such parameters are system size, local dimension, temperature,
density of states, etc~\footnote{We shall remark that there may not be a sharp distinction
between tunable (or controllable) parameters and untunable ones.
Such a boundary can depend on technology, for instance.}.

We find that one has to further relax the definition of universality.
Generically, universality means that the group $SU(d)$ for any $d$ can be realized efficiently
to \emph{arbitrary} accuracy based on a universal gate set.
The Solovay-Kitaev algorithm~\cite{DN06}
ensures that arbitrary accuracy can be achieved efficiently.
We realize that such a definition of universality is strong
in the sense that it allows arbitrarily long accurate quantum computation.
However, in practice a computation is not arbitrarily long.
As a result, we introduce a weak version, coined
`quasi-exact' universality,
which is suitable for computation of finite length and limited accuracy;
however, the accuracy is required to be tunable,
hence improving the ability to perform accurate computation of any desired length
and eventually approaching universality.

The quasi-exact universality
can be supported on quasi-exact codes.
In the quasi-exact scenario,
each logical gate has a finite accuracy
and there are in total a finite number of distinct logical gates.
There is no infinitesimal logical gate, in particular,
which would be treated as a non-perfect logical identity gate.
Furthermore, we show that there exist quasi-exact codes such that
when tuning the scaling parameters,
the accuracy parameter becomes smaller and
the number of transversal logical gates increases,
approaching universality for a certain unitary group.
This is in contrast to the exact setting
for which the size of the transversal logical gate set is fixed.


In this work we find a class of quasi-exact codes which support
a set of transversal logical gates for $SU(d)$.
This class of codes are from $SU(d)$ generalizations of
valence-bond solid (VBS) models~\cite{AKLT87,GR07,KHK08,MUM+14},
which appear as ground states of 1D local Hamiltonians with symmetry-protected topological (SPT) orders~\cite{CGW11,SPC11,CGL+13,DQ13a,DQ13b}
and have proven to be useful in the measurement-based
model~\cite{RB01,GE07,WSR17,SWP+17}.
This class of codes are Lie-group covariant codes,
the power (and limitation) of which have been demonstrated recently~\cite{Pre00,HNP+17,FCS+18,FNA+19,WA19}.
The SPT order enables covariance and quasi-exact universality,
and the local errors we consider can be erasure or arbitrary ones on a local site.
The number of logical gates scales as $Nd_q$,
the inverse of the accuracy of the code,
for $N$ as the system size, $d_q$ as the local dimension.
We believe better scaling can be achieved for other types of quasi codes to be found.
Our work demonstrates the possibility of fault-tolerant quantum computation with
(quasi-)universality and transversality.


\section{Quasi codes}\label{sec:qcode}

We first recall the setting of exact codes.
An exact encoding can be defined by an isometry
$ V: \C H_L \ra \C H $
from the \emph{logical} space $\C H_L$ to the \emph{physical} space $\C H$
such that $V^\dagger V=\I$ and $VV^\dagger=P$,
for $P$ as the projector onto the code space $\C C \subset \C H$.
The code dimension is denoted as $d_L:=\text{dim}\C H_L=\text{dim}\C C$,
and the physical dimension is $d_Q:=\text{dim}\C H$.
When $\C H=\otimes_{n=1}^N \C H_n$
for a system with $N$ local sites $Q_n$ each of dimension $d_{q,n}$,
$d_Q=\prod_n d_{q,n}$.

The correctability condition~\cite{KL97} for a channel $\C N$
defined by a set of error operators $\{E_i\}$ on $\m{H}$ is
\be  P E_i^\dagger E_j  P = a_{ij}  P, \label{eq:qec}\ee
for $[a_{ij}]$ as a density operator.
The recovery channel $\C R=\{R_k\}$ is found by diagonalizing $[a_{ij}]$ to be $[d_k]$,
and define \be R_k:= P F^\dagger_k/\sqrt{d_k} \label{eq:qeck} \ee for $d_{k}\neq 0$
and $\{F_k\}$ as the rotated representation of the noise channel $\C N$.
Meanwhile, the condition for detection is
$  P E_i  P = e_i  P,\; e_i \in \B C.$
Each error operator $E_i$ could be local or nonlocal.
When $\{E_i\}$ is a local basis of the space of bounded operators
on a local site $\C B(\C H_n)$
(the subscript $n$ is omitted for $E_i$) for any $n$,
then for any $F_j \in \C B(\C H_n)$, 
condition~(\ref{eq:qec}) 
implies $  P F_j^\dagger F_l  P \propto  P$, 
which means arbitrary local error can be corrected.

Conditions for approximate error correction have been established,
and optimizations are needed to find a recovery scheme~\cite{RW05,BO10,BO11,Ben11,NM10,MN12}, in general.
It holds that
\be  P E^\dagger_i E_j  P = a_{ij}  P +  P B_{ij} P,  \label{eq:qeqec}\ee
for $[a_{ij}]$ as a density operator
and operators $B_{ij}$ as uncorrectable parts.
The exact case~(\ref{eq:qec}) is for vanishing $B_{ij}$.
The recovery error by a channel $\C R$
can be measured by the diamond-norm distance~\cite{KSV02}
$d_\diamond(\C Q, \I):=\| \C Q-\I \|_\diamond$,
for $\C Q:=\C V^\dagger \C R \C N \C V$,
$\C V(\cdot):=V(\cdot)V^\dagger$ as the encoding map,
which is equivalent to the trace distance $D_t(\C C_{\C Q}, \omega)$ by~\cite{Wat18}
\be \frac{1}{2d_L}d_\diamond(\C Q, \I)\leq D_t(\C C_{\C Q}, \omega) \leq \frac{1}{2}d_\diamond(\C Q, \I), \label{eq:inf}\ee
for $\omega:=|\omega\ket\bra\omega|$,
$|\omega\ket:=\sum_{i=1}^{d_L}|ii\ket/\sqrt{d_L}$ as the maximally entangled state,
and $\C C_{\C Q}:=\C Q\otimes \I (\omega)$, known as a Choi matrix.
Other measures can also be employed such as entanglement fidelity~\cite{RW05,GLN05,Sch96}
(also see Appendix~\ref{sec:appqec}).
We say an approximate code is $\epsilon$-correctable for a noise channel $\C N$
when there exists a recovery channel $\C R$ such that
$d_\diamond \leq \epsilon$~\cite{BO10,BO11,Ben11,NM10,MN12}.

Now we introduce quasi-exact codes.
We will use `quasi' as a shorthand for `quasi-exact.'
As explained in the Introduction,
the motivation is to make the recovery error as small as possible,
and the idea is to introduce in some scaling parameters $\vec{\lambda}$
such that the recovery error becomes tunable.
The recovery channel $\C R$ may depend on $\vec{\lambda}$ in principle;
however, in the vicinity of exact codes,
we expect that the recovery scheme~(\ref{eq:qeck}) can be employed.
In accordance with the notion of quasi universality to be defined later,
we defined quasi codes as follows.

\begin{definition}[Quasi codes]
A quasi code $\C C(\vec{\lambda})$ is a family of
$\epsilon(\vec{\lambda})$-correctable approximate codes with the recovery scheme~(\ref{eq:qeck}),
such that each code is defined with fixed values of $\vec{\lambda}$,
which is a vector of a finite number of real scaling parameters,
and $\epsilon(\vec{\lambda})\ra 0$
at some points in the parameter space of $\vec{\lambda}$.
We assume that $\epsilon$ is a smooth function of $\vec{\lambda}$.
\end{definition}

We see that the quasi setting is a particular and strengthened case of the approximate setting.
We shall remark on a few points.
i) We assume the number of scaling parameters is finite.
ii) A quasi-to-exact (QTE) limit is a limit by tuning some parameters
of $\vec{\lambda}$ so that $\epsilon(\vec{\lambda}) \ra 0$ is achieved.
Each limiting (`critical') point is an exact code.
iii) One might demand $\epsilon$ be a monotonic function of each parameter of $\vec{\lambda}$,
while this is not required by definition.

There exists an explicit form of the recovery error in terms of
the trace distance $D_t(\C C_{\C Q}, \omega)$~(\ref{eq:inf}).
We find
$  \C Q (\hat{\sigma}) = \hat{\sigma}+ \sum_{kl}\hat{B}_{kl}
 \hat{\sigma} \hat{B}_{kl}^\dagger /d_k  $
for any logical state $\hat{\sigma}$, $\hat{B}_{kl}:=V^\dagger B_{kl} V$.
This means the logical state is recovered up to the
perturbation by the set $\{ \frac{1}{\sqrt{d_k}}\hat{B}_{kl} \}$.
We obtain the trace distance
\be D_t(\C C_{\C Q}, \omega)=\frac{1}{2d_L}\sum_{kl} \frac{1}{d_k}\T tr (\hat{B}_{kl}^\dagger \hat{B}_{kl}), \label{eq:Mch} \ee
which shall be a smooth function of each $d_k$, $\beta_{kl}:=\T tr (\hat{B}_{kl}^\dagger \hat{B}_{kl})$,
and the cardinality of the set $\{k\}$, i.e., the size of the `environment', $d_E$.
For local-error models,
if the system size is $N$ and local site dimension is $d_q$,
then $d_E=\text{poly}(N, d_q)$; e.g.,
$N^{c_1}d_q^{c_2}$ to the leading order for some positive constants $c_1$, $c_2$.
The variables $\beta_{kl}$ and $d_k$ should show universal decay behaviors with some scaling parameters
such as $N$ and $d_q$ in order to make $D_t$ small.
If $D_t \ra 0$, hence $d_\diamond \ra 0$, under a QTE limit, it is quasi correctable.
This also implies that for bounded operators in the span of $\{E_i\}$,
the $D_t$ in terms of the new set of $\beta_{kl}'$ and $d_k'$ should also decay in a QTE limit,
making it quasi correctable.

\section{Quasi universality}
\label{sec:qcompt}

Now we introduce the notion of quasi universality,
which is suitable for computing with quasi codes,
and can also be employed in other settings.
Universality means that on a logical space of dimension $d$,
the whole group $SU(d)$ can be
realized efficiently~\cite{puncture}.
The Solovay-Kitaev algorithm~\cite{DN06} proves that,
given any $U\in SU(d)$,
there exists an efficient classical algorithm 
that produces an efficient sequence of logical gates $U'=\prod_i U_i$
with the operator-norm distance $d(U,U')$ arbitrarily small.

In the setting of universality above,
the logical identity gate $\I$ is not explicitly written but implied.
However, there is no perfect $\I$ for quasi codes
due to the error correction features (see Eq.~(\ref{eq:Mch})).
The leftover from error correction is random,
which means that the logical identity gate $\I$ shall be replaced by a set of gates
that are close to it, with a proper distance measure
such as entanglement fidelity,
trace distance, or operator norm.
Denote this set as $\mathfrak{I}_\eta$ for $\eta$ as a measure of distance from $\I$.
For a quasi code, $\eta$ will be its accuracy $\epsilon(\vec{\lambda})$.
The set $\mathfrak{I}_\eta$ is treated as an equivalence class of gates,
and any gate $\I_\eta \in \mathfrak{I}_\eta$ is treated the same as $\I$.
We call such a set a `gate-cell', or just `cell',
and $\eta$ as the `size' or accuracy of it.
Clearly it also induces gate-cells for generic logical gates.
A gate-cell $\mathfrak{U}_\eta$ of accuracy $\eta$ for a logical gate $U\in SU(d)$ is a dense set of gates
such that each $U_\eta \in \mathfrak{U}_\eta$ is of distance $\eta$ from $U$.
Therefore, the whole group $SU(d)$ can be partitioned into a collection of
non-overlapping gate-cells of various sizes.
One could pick a proper representative gate for each gate-cell,
and denote the set of them as the coarse-grained set $SU(d)_\eta$,
and it is clear that $SU(d)_\eta \subset SU(d)$.
For any $U \in SU(d)$,
there exists an element in $SU(d)_\eta$ of distance no more than $\eta$ from $U$.
The partition into cells can be non-uniquely chosen and shall be kept fixed
in order to define distinct logical gates.

\begin{definition}
[Quasi universality]
A computation on a quasi code $\C C(\vec{\lambda})$ with accuracy $\epsilon(\vec{\lambda})$
is quasi universal for a unitary group $SU(d)$ if
the coarse-grained set $SU(d)_{\epsilon(\vec{\lambda})}$ can be realized efficiently,
and the group $SU(d)$ can be approached when $\epsilon(\vec{\lambda})\rightarrow 0$ in a certain QTE limit.
\end{definition}

Generically, for quasi universality the number of distinct logical gates is finite,
which yet is quasi infinite due to the ability to improve the accuracy of a quasi code.
The transition from quasi universality to universality is not required to be exponentially fast
(with respect to some scaling parameters),
which is desirable yet not necessary.
Different transition behaviors as functions of scaling parameters would lead to a
classification of quasi codes that are quasi universal for a certain unitary group.

Arbitrarily long computation cannot be reliably done with quasi codes
since the leftover from error correction can accumulate.
Let $\eta$ denote the accuracy $\epsilon(\vec{\lambda})$ for simplicity.
Suppose we need to approximate $U\in SU(d)$ to accuracy $\varpi$,
what would be the relation between $\varpi$ and $\eta$?
For quasi codes that any gate $U$ can be realized directly,
$\varpi$ is simply $\eta$.
For the setting that each gate $U$ is built from a product of gates
from a universal gate set, denoted as $\C S$,
$\varpi$ is lower bounded by $\eta$.
Namely, for $U_\eta=U \I_\eta$, $V_\eta=V \I_\eta$,
for $\I_\eta\in \mathfrak{I}_\eta$,
the distance $d(U_\eta V_\eta,UV)\leq 2\eta$
if $d(U_\eta,U)\leq \eta$, $d(V_\eta,V)\leq \eta$~\cite{cell}.
In general, given any $U\in SU(d)$,
first use gate-synthesis algorithms~\cite{DN06}
to approximate $U$ by $U'=\prod_L \bigotimes_J U_{LJ}$,
for $U_{LJ}$ as gates from a universal gate set $\C S$,
$L$ as the label of layers relating to the time of computation,
and $J$ as the label of gates in each layer relating to the size of computation.
Denote the distance $d(U,U')$ as $\varpi_0$, which could be as small as zero.
As each $U_{LJ}$ will be approximated by $U_{LJ,\eta}$,
we see that $d(U',U'_\eta)\leq m\eta$ for $m:=\sum_L |J|_L$ as the number of gates.
Then the distance $d(U,U'_\eta)\leq m\eta+\varpi_0$.
As a result, we could obtain a lower bound for the accuracy
\be \varpi \geq m\eta, \ee
which is also the upper bound on the size $m\leq \frac{\varpi}{\eta}$.
For the VBS covariant codes that we find below,
$\eta$ scales as the inverse of the given system size and local dimension,
which provides a linear size constraint on the computation.

\section{Transversal logical gates}

Now we restrict logical gates to be transversal,
which is a primary scheme to achieve fault-tolerance.
However, it is known that transversality is incompatible with
universality~\cite{EK09} (see Appendix~\ref{sec:eknogo}),
and this also extends to approximate codes~\cite{FNA+19,WA19}.
With the quasi universality defined above,
which is weaker than the requirement of universality,
here we argue that transversality is compatible with quasi universality,
and there are quasi codes that can achieve both.

We first analyze how quasi error correction modifies
the original group-theoretic argument~\cite{EK09}.
Transversal gates are of the form
\be U=\bigotimes_j U_j \label{eq:trans}\ee
for $j$ as the index of subsystems.
The set of transversal logical gates is a Lie group,
denoted as $\C G$,
and the connected component of identity $\C G^0 \subset \C G$ is logically trivial for exact codes~\cite{EK09}.
We will show now for quasi codes this may not be the case.
Let $e^{i\xi D}\in \C G^0$, $\xi\in \mathbb{R}$ then $D=\sum_j \alpha_j H_j$
for $H_j$ acting on the subsystem $j$, and $DP=PDP$.
Given the error set $\{E_{jl}\}$ on subsystem $j$,
$H_j$ can be written as $H_j=\sum_l \beta_{jl}E_{jl}$.
As $PE_{jl}P=e_{jl}P+PB_{jl}P$ from quasi error correction,
then \be DP=hP+PBP\ee
for $h:=\sum_{jl}\alpha_j\beta_{jl}e_{jl}$,
$B:=\sum_{jl}\alpha_j\beta_{jl}B_{jl}$.
The terms $D^n P$ can be computed, and we find
$D^n P \approx \sum_{m=0}^n h^m P(BP)^{n-m}$.
Thus, \be e^{i\xi D} P \approx e^{i\xi h} e^{i\xi PBP} P. \ee
The unitary operator $e^{i\xi PBP}$
is the term due to quasi error correction.
If $\xi\in o(1)$, then $e^{i\xi D}$ falls in the cell of logical identity.
If $\xi \in O(1)$, then $e^{i\xi D} $ can have a nontrivial logical action.
This means that
there exist quasi codes for which
the connected component of identity $\C G^0 \subset \C G$
does not collapse to logical identity.
Instead, $\C G^0$ will permit a gate-cell structure and split into a collection of cells of various sizes.
The analysis can also be extended to subsystem quasi codes (Appendix~\ref{sec:sub}).

The gate-cell structure of $\C G^0$
may appear surprising since a tiny recovery error shall not affect the logical structure of $\C G^0$
significantly.
The reason is that the recovery error is a measure of the average effects of local noise operators,
in terms of a noise channel,
while logical gates are from the accumulated net effects of local operators.
A logical gate on a quasi code shall have a weight (i.e., support) large enough
compared with the system size.



As the group $\C G$ is compact for a finite system size,
the number of logical gates would be finite.
This is \emph{not} a problem for quasi universality
since an infinite number of logical gates is not required.
Furthermore, when a quasi code tends to be exact
in a QTE limit,
the sizes of gate-cells become smaller,
and the number of logical operators could tend to be infinite.
This is distinct from any exact quantum code
for which the number of logical gates is fixed.
This demonstrates that
there exist quasi codes that can achieve transversality and quasi universality.
We will see below that the $SU(d)$-covariant quasi codes achieve this for $SU(d)$,
while the sizes of gate-cell for all logical gates are of the same order.

\section{$SU(d)$-covariant quasi codes}\label{sec:vbsc}

We find classes of covariant codes can be defined from matrix-product states~\cite{PVW+07,Sch11}
with global continuous symmetry, namely,
the higher-symmetry generalizations of VBS states~\cite{AKLT87,GR07,KHK08,MUM+14},
which have SPT order and appear as ground states of 1D local Hamiltonians.
Below we define covariant codes with unitary symmetry,
while the construction can be easily generalized to other Lie groups such as orthogonal
and symplectic groups, and also to higher spatial dimensions.

The $SU(d)$-covariant codes we consider are defined by the isometry
$V:|\alpha\ket\mapsto |\psi_\alpha\ket$ for $|\alpha\ket\in \C H_L$ and
\begin{equation}\label{}
  |\psi_\alpha\rangle=\sum_{i_1,\dots,i_N=0}^{d^2-1} A^{i_N}_N\cdots A^{i_1}_1 |\alpha\rangle
  |i_1\rangle \cdots |i_N\rangle \in \C C,
\end{equation}
with the logical space $\C H_L$ as the fundamental representation of $SU(d)$,
and the physical space $\C H=\C H_L \otimes (\otimes_{n=1}^N \C H_n)$,
and each $\C H_n$ as the adjoint representation of the group $SU(d)$.
The $N$ sites $\C H_n$ are known as `bulk', and the site $\C H_L$ as `edge.'
Both the edge and bulk states are employed for the codes.
The tensors $A^{i_n}_n$  ($n=1,\dots,N$) are translation-invariant
(hence $n$ can be omitted) taking the form $A^i=\sqrt{\frac{2d}{d^2-1}} t^i$ for Gell-Mann matrices $t^i$ with
$\T tr(t^it^j)=\frac{1}{2}\delta_{ij}$,
$[t^i,t^j]=if_{ijk}t^k$ for $f_{ijk}$ as structure constants of $SU(d)$.
The tensors $A^i$, also called Kraus operators,
form a quantum channel $\mathcal{E}$ with
$\sum_{i} A^{i\dagger} A^i=\mathds{1}$.
The symmetry is
\begin{equation}\label{eq:symcond}
  \sum_j u_{ij}(g) A^j=\breve{U}(g) A^i \breve{U}(g)^\dagger
\end{equation}
for $U(g)=[u_{ij}](g)$ of size $d^2-1$, $\breve{U}(g)$ of size $d$,
for $g\in SU(d)$~\cite{CGW11,SPC11}.
The channel $\mathcal{E}$ can be dilated to a unitary operator $W$ such that
$W |0\rangle=\sum_{i} A^i |i\rangle$, which is an isometry.
The encoding isometry $V$ is from the product of $W$, each as the dilation of $\mathcal{E}$.
From the global symmetry, it is clear that
\be V \breve{U}(g)|\alpha\ket= U(g)\otimes U(g)\otimes \cdots \otimes U(g) \otimes \breve{U}(g)  |\psi_\alpha\rangle \ee
for $N$ factors of $U(g)$ acting on the bulk and $\breve{U}(g)$ acting on the edge, for $g\in SU(d)$.

Now we study error correction property of the codes.
We label the edge as the ($N$+1)th site, and others from 1 to $N$ sequentially.
The edge before encoding can be viewed as the 0th site.
The local state of the edge is
\be \sigma_{N+1}= \C E^N(|\alpha\ket\bra \alpha|)= \mathds{1}/d +
2\chi^N \sum_a t^a t^a_{\alpha\alpha}\ee
for $\chi:=\frac{-1}{d^2-1}$, $t^a_{\alpha\beta}:=\bra \alpha|t^a |\beta\ket$.
This means that partial information of the logical state can be read off from
the edge state via $t^a_{\alpha\alpha}$,
yet this is exponentially suppressed when $N$ increases,
and $\sigma_{N+1} \ra \mathds{1}/d$, which is the unique fixed point of $\C E$.
The channel $\C E$ has other eigenvalues as $\C E (t^a)=\chi t^a$.
The local state $\rho_n$ ($n\in [1,N]$) of the $n$th bulk site is
\be  \rho_n=\sum_{i j}  \T tr [  \sigma_{n} A^jA^i  ] |i\ket\bra j|, \ee
which is the complementary state of
$ \sigma_{n}:=\C E^{n-1}(|\alpha\ket\bra \alpha|)$.
It is easy to see $\rho_n$ converges to the completely mixed state exponentially,
$\rho_n \ra \mathds{1}/(d^2-1)$ as $n$ increases,
yet for small $n$, the local state $\rho_n$
contains the observable $t^a_{\alpha\alpha}$ of the logical state $|\alpha\ket$.
This manifests that the codes are indeed quasi codes.

To define a proper error model,
observe that the action of a local operator $T^a$,
as the adjoint representation of a Gell-Mann matrix $t^a$,
can be converted to actions on the edge.
The action of $T^a$ on a local site is
$T^a\sum_i A^i |i\ket=\sum_i [A^i,t^a] |i\ket$.
Denote the link $(n,n\pm 1)$ as $n\pm$,
then the action of $T_n^a$ on a local site $n$
is a superposition of the actions of
$t_{n+}^a$ and  $t_{n-}^a$ on the edge space.
So we could view the information encoded in the bulk equivalently
as encoded in the links, i.e., the history states of the edge.

For the local error set $\{t_{n+}^a\}$, we find
\be \bra \psi_\alpha|  t_{n+}^a  |\psi_\beta\ket = \chi^n t^a_{\alpha\beta}, \label{eq:qde}\ee
and also
$\bra \psi_\alpha|  t_{n-}^a  |\psi_\beta\ket = \chi^{n-1} t^a_{\alpha\beta}$.
This means local errors are only approximately detected.
When $d\ra \infty$, or $n \ra \infty$, the detection becomes exact.
For correlation functions we find
\be \bra \psi_\alpha|  t_{m+}^a t_{n+}^b  |\psi_\beta\ket = \chi^{n-m} \delta_{ab}\delta_{\alpha\beta}/2d+\chi^nh_{bac} t^c_{\alpha\beta}/2, \label{eq:qco} \ee
for $n>m\geq 0$, $h_{bac}=d_{bac}+if_{bac}$,
and $d_{bac}$ are also structure constants of $SU(d)$.
We see that this correlation function contains information $t^c_{\alpha\beta}$ of the logical states,
which is suppressed for large $n$ or large $d$.
In addition, this leads to
$\bra \psi_\alpha| T_n^a t_{N+1}^b |\psi_\beta\ket = \frac{-d\chi^{N-n}}{2(d^2-1)} \delta_{\alpha\beta} \delta_{ab}$,
$\bra \psi_\alpha|  T_m^a T_n^b  |\psi_\beta\ket = \frac{-d^3\chi^{n-m-1}}{2(d^2-1)^2} \delta_{\alpha\beta} \delta_{ab}$.
We see that the edge is more correlated to bulk sites that are close to it,
and there are exponential decay of bulk correlation functions.



The quasi conditions (\ref{eq:qde}) and (\ref{eq:qco}) also apply
to errors in the span of $\{t_{n+}^a\}$ with $t^0\equiv \mathds{1}$.
It is not hard to see that a logical error gate resulting from an error on the link $n+$
depends on $n$, and can be approximated as a unitary operator
$E_n=e^{i\chi^n \sum_k  \epsilon_k t^k}$, $\epsilon_k\in \mathbb{R}$.
We shall average over random errors that occur on the system,
and the net effect is a random unitary channel
\be \C N (\sigma)=\frac{1}{N}\sum_n E_n \sigma E_n^\dagger,\; \forall \sigma \in \C B(\C C), \ee
which, to the order $O(\chi^{2n})$ for each $n$, can be approximated by a unitary operator
$e^{i\eta \sum_k \epsilon_k t^k}$
for \be \eta:=\frac{\chi}{N}\frac{1-\chi^N}{1-\chi}, \label{eq:eta}\ee
which vanishes when $N\ra \infty$ or $d\ra \infty$
as two types of QTE limits.

The error analysis above is consistent with the erasure error model,
for which the performance of covariant codes has been studied~\cite{FNA+19}.
From (\ref{eq:qde}), we find
$\bra \psi_\alpha|  T_n^a  |\psi_\beta\ket = \frac{d^2\chi^{n-1}}{d^2-1}   t^a_{\alpha\beta}$,
$ \bra \psi_\alpha|  t_{N+1}^a  |\psi_\beta\ket = \chi^N t^a_{\alpha\beta}$,
and then $t^a_{\alpha\beta}= \bra \psi_\alpha|
t_{N+1}^a  |\psi_\beta\ket+\sum_n \bra \psi_\alpha|  T_n^a  |\psi_\beta\ket$.
If the environment can erase a local site $n\in [1,N+1]$ randomly,
then it can read out the logical value $t^a_{\alpha\beta}$ from a global observable.
Then the uncorrectable part of the code is lower bounded by a quantity
proportional to $(N \max_n \Delta T_n)^{-1}$,
for $\Delta T_n$ as the spectral range of a local observable $T_n$,
and $\max_n \Delta T_n$ scales with the dimension $d^2-1$,
which agrees with the scaling of $\eta$~(\ref{eq:eta}).

When the codes are prepared as ground states of frustration-free
local Hamiltonians~\cite{AKLT87,GR07,KHK08,MUM+14},
which in general take the form
$H=\sum_n h_{n,n+1}$,
the nearest-neighbor interaction terms $h_{n,n+1}$, although do not commute with each other,
play similar roles with stabilizers~\cite{WAR18}.
Each term $h_{n,n+1}$ shall be minimized for the code space $\C C$.
An error $E_n$ will increase the energy term $h_{n,n+1}$,
and can be corrected by cooling back to $\C C$.

For quantum computing with error correction,
the sequence of gate operations is $\prod_{\ell=1}^L U_\ell E_\ell$
with the enacted gates $U_\ell$
interrupted by error gates $E_\ell$.
Product of $U_\ell$ can yield the whole group $SU(d)$.
Recall that $E_\ell$ depends on $\eta$,
and when $d\ra \infty$, or $N\ra \infty$, then $E_\ell\ra \mathds{1}$.
For finite $d$ and $N$, when the length $L$ of the computation increases,
the errors will increase, too.
As we already know, the quasi feature limits the accuracy of logical gates.
The parameter $\eta$ can be viewed as the unique measure of the accuracy of logical gates,
although each error gate $E_\ell$ can be different in practice.
Therefore, any logical gate is actually a gate-cell of size measured by $\eta$,
and gates within this accuracy are equivalent.
A $SU(d)$-covariant code defined above is quasi universal for $SU(d)$,
and it becomes exact universal for $SU(d)$ when the code itself approaches exact.

\section{Conclusion}\label{sec:conc}

In this work, we have studied the model of quasi-exact quantum computation with two central concepts:
quasi codes and quasi universality.
The valence-bond solids codes we find can provide a quasi-continuous universal and transversal set of logical gates,
with the number of logical gates proportional to the product of the system size and local dimension.
We remark that by relaxing the requirement of covariance and allowing various sizes of gate-cells,
it will naturally lead to the setting that a discrete universal gate set is firstly realized,
and all other gates are from product of them.
It is desirable to find a quasi code that can satisfy the quasi universality,
transversality, and discreteness of gate set simultaneously.


A slightly easier task is to employ concatenation of codes such that their transversal sets of
gates can be combined to be universal.
The concatenation may also benefit the allowed error threshold.
A quasi code can be taken as either an inner code or an outer code,
and concatenated with another quasi code or exact code, such as stabilizer codes.
It is thus also important to see if there are quasi codes that permit quasi universality
with such concatenated transversality.

\section{Acknowledgement}

This research was supported
by the Government of Ontario and the Government of
Canada through ISED.
We thank an anonymous referee for valuable comments.
Discussions with A. M. Alhambra, S. Bravyi, A. W. Cross, P. Faist, A. Kubica, M. Vasmer, T. Yoder, T. Lan, and Y. Wang are acknowledged

\appendix


\setcounter{figure}{0}
\setcounter{table}{0}
\setcounter{equation}{0}
\setcounter{section}{0}
\setcounter{page}{1}


\renewcommand{\theequation}{A\arabic{equation}}
\renewcommand{\thefigure}{A\arabic{figure}}
\renewcommand{\thetable}{A\arabic{table}}

\section*{Appendix}

\subsection{Approximate quantum error correction}
\label{sec:appqec}

According to~\cite{BO10}, a channel $\C N$ on a code $P$ is $\epsilon$-correctable iff
\be  P E^\dagger_i E_j  P = a_{ij}  P +  P B_{ij}  P ,  \label{eq:bo}\ee
for $ d(\C D+\C B, \C D) \leq \epsilon$,
with $\C D (\rho) = \rho_*$,
$\C B (\rho) =\sum_{ij} \T tr(\rho B_{ij}) |i\ket\bra j|$,
$\rho\in \C C$,
for $\rho_*:=\sum_{ij}a_{ij}|i\ket\bra j|$, $B_{ij}\in \C B(\C H)$,
$d$ as a proper distance measure on channels,
e.g., the diamond norm distance $d_{\diamond}$~\cite{KSV02},
trace distance on Choi states, or
Bures distance $d_{B}=\sqrt{1-F}$ for the entanglement fidelity $F$~\cite{Sch96}.
The channels $\C D$ and $\C B$ maps system states to `environment' states
since $\{|i\ket\}$ are states of an environment, $E$.
Furthermore, there exists a recovery channel $\C R$ such that $d(\C V^\dagger \C R \C N \C V,\I)\leq \epsilon$.
The existence of a near-optimal recovery channel is also proved,
but it may require a numerical convex optimization to find it~\cite{RW05,BO10}.

The $\epsilon$-correctability of the set $\{E_i\}$
does not guarantee that for other operators in its span, however.
Namely, suppose another set of error operators $\{F_\ell\}$ is defined by a matrix
$\Upsilon=[y_{\ell i}]$
such that $F_\ell=\sum_i y_{\ell i} E_i$.
Then the superoperator $\Upsilon$ defined by conjugation of $\Upsilon$ induces two new channels
$\tilde{\C D}=\Upsilon \circ \C D$ and $\tilde{\C B}=\Upsilon \circ \C B$
to the condition~(\ref{eq:bo}) for the new set $\{F_\ell\}$.
As $\Upsilon$ is not contractive in general, namely, it does not necessarily reduce distance between states,
the distance $d(\tilde{\C D}+\tilde{\C B}, \tilde{\C D})$ is not upper bounded by $\epsilon$ anymore.
This poses a problem when considering the approximate correction of arbitrary local errors,
and instead motivates the introduction of quasi codes
which come with a set of scaling parameters.

\subsection{Eastin-Knill no-go theorem}
\label{sec:eknogo}

We review the contents of Eastin-Knill no-go theorem~\cite{EK09},
which basically states that there is no universal set of
transversal logical gates (TLG) supported by a finite-dimensional physical system.

A unitary operator $U$ is a logical operator iff \be UP=PUP. \label{eq:logg}\ee
A unitary operator $U^\dagger$ is a logical operator iff $PU=PUP$.
It turns out, if $U$ is logical, then $U^\dagger$ is also logical.
This means a state in the code space cannot be mapped out of it by $U$,
while a state out of the code space cannot be mapped into it by $U$, neither.
It is also equivalent to say $[U,P]=0$.

In order to define transversal gates,
we need to introduce `transversal part', or subsystem.
  An error-correction subsystem, or subsystem for short,
  is a part of the whole system for which error correction can be performed.
The natural choice of a subsystem is a local site.
However, a subsystem can consist of several local sites,
which should usually be a connected local part of the whole system.
This applies to topological codes that have macroscopic code distances.

A TLG does not couple different non-overlapping subsystems within the same logical qubit,
and only couples a corresponding subsystem from different logical qubits.
TLG do not spread out errors across subsystems for the same logical qubit.
So the error-correction for each logical qubit can ensure fault tolerance.

For many logical qubits each encoded by a different physical system, denoted as $Q[r]$,
usually a one-to-one correspondence of subsystems has to be chosen.
For instance, for two logical qubits $Q[r]=\bigcup_n Q_n[r]$, $r=1,2$,
a subsystem $Q_n[1]$ can be chosen to correspond to $Q_n[2]$ for the same label $n$.
In general, there might be a permutation $\pi(n)$.
If each system comes with a Hamiltonian $H[r]$,
then the total Hamiltonian is the sum of them $H=\sum_r H[r]$ without interaction terms.

A transversal gate takes the form
\be U=\bigotimes_j U_j \label{eq:trang}\ee
for $j$ as the index of subsystems.
Note that $U$ is defined up to any permutation of subsystems
since permutation cannot spread out errors except the locations of them.


In order to see the generality of Eastin-Knill theorem, below we review its content in details.
There are several crucial assumptions.
(a) The error detection is exact for an error set $\{E_i\}$,
which spans each local subsystem.
(b) Transversality is fixed, namely, all TLG takes the form~(\ref{eq:trang}) for a given partition of subsystems.
(c) The Hilbert space dimension of the system is finite.
(d) The universality is exact, i.e., the group $SU(d)$ shall be realized efficiently
to arbitrary accuracy.

Given a code space $P$, it first shows that the set of logical gates~(\ref{eq:logg})
form a Lie group $\C L$.
Given a transversality and the form~(\ref{eq:trang}),
it shows that the set of TLG is also a Lie group $\C G=\C L\bigcap \C A$,
for $\C A=\bigotimes_j U(d_j)$, $d_j$ as the dimension of a subsystem.
Now the connected component of identity $\C G^0$ in $\C G$ contains elements of the form
$C=\prod_k e^{i\xi_k D_k}$, $\xi_k\in \mathbb{R}$.
An operator $e^{i\xi D}$ is a TLG for $\xi\in \mathbb{R}$,
and then it shows $DP=PDP$.
The operator $D$ can be written as a sum of local terms
$D=\sum_j \alpha_j H_j$ due to the structure of Lie algebra,
and each $H_j$ acts on the subsystem $j$.
Given the detection of arbitrary local errors,
it holds $PH_jP\propto P$, and then
$DP=PDP\propto P$.
As the result,
$CP\propto P$,
which means that $C$ acts as the logical identity gate,
and the whole group $\C G^0$ `collapses' to identity.
As the quotient group $\C Q=\C G/\C G^0$ is a topologically discrete group,
the number of logically distinct operators is finite.
In other words, the set of TLG is not universal.

Next we remark on some points.
(1) During the execution of each logical gate,
there may be leakage out of the code space,
as long as it goes back to the code space at the end.
(2) Each local unitary $U_j$ can be realized in many ways, even not unitarily,
as long as the net effect is unitary.
For instance, ancilla and measurement can be used.
(3) There is no logical ancilla to realize a logical gate $U$
since $U$ itself must be unitary of the form~(\ref{eq:trang}).
(4) A subsystem can contain several local sites,
and this especially applies to codes with large code distance,
such as topological codes.
Error correction on local sites ensures error correction on a subsystem.
For code distance $d=2t+1$,
a subsystem can be as big as $t$.
This means that each $U_j$ can be an entangling gate on the underlying local sites.
However, the transversality has to be fixed to ensure that
all logical gates are of the form~(\ref{eq:trang}).
(5) The theorem applies to arbitrarily large but finite dimensional Hilbert space.
This fact is crucial to generalize it to the quasi-exact setting, e.g.,
the group-theoretic argument still applies.

\subsection{Subsystem quasi codes}
\label{sec:sub}

Here we extend the results to the subsystem quasi codes,
which are the quasi version of subsystem codes.
In this case, the code space $\C C$ has a tensor product form
\be \C C= \C T \otimes \C J\ee
for the truly logical subspace $\C T$ that encodes the information
and a junk `gauge subspace' $\C J$ that does not encode logical information.
The apparent encoding operator is now not one-to-one
$V:|i\ket\mapsto |\psi_i\ket |\psi_j\ket$, for
$|i\ket\in \C H_L$, $|\psi_i\ket\in \C T$,
and any state $|\psi_j\ket\in \C J$ (which can also be mixed states),
but it is effectively one-to-one when the $\C J$ part is ignored.
The total space takes the form
\be \C H \cong \C T \otimes \C J \oplus \C S,\ee
for $\C S$ as the syndrome subspace.
For error correction, in additional to leakage to $\C S$,
errors may also generate entanglement between  $\C T$ and $\C J$.
A good subsystem code is designed such that $\C J$ can benefit the error correction.
We use $P$, $P_{\C T}$, and $P_{\C J}$ as the projector on $\C C$,
$\C T$, and $\C J$, respectively.
We assume an orthonormal basis of $\C J$ can be chosen to define $P_{\C J}$.

The exact error correction for a set of errors $\{E_i\}$ on subsystem codes
has been shown~\cite{KLP05,Pou05,NP07} to be
\be P E_i^\dagger E_j P = (\I_{\C T} \otimes J_{ij} )P,\ee
for $\I_{\C T}$ ($J_{ij}$) acting on the space $\C T$ ($\C J$).
Contrary to the standard condition~(\ref{eq:qec}),
here there can be a nontrivial action $J_{ij}$ on the gauge part.
The condition above is equivalent to
\be P_{\C T} E_{ig}^\dagger E_{jh} P_{\C T} = a_{ijgh} P_{\C T},\ee
for $E_{ig}:=E_i |\psi_g\ket$, with any $|\psi_g\ket\in \C J$.
The effective error operators $E_{ig}$ are rectangular,
namely, they can map states in $\C T$ to states in $\C T$ and $\C J$.
The span of $\{E_{ig}\}$ is the product of the span of $\{E_i\}$ and $\{|\psi_g\ket\}$.
If the dimension of $\C J$ is $d_{\C J}$,
then a set of $d_{\C J}^2$ linearly independent states $\{|\psi_g\ket\}$
is enough to ensure the correction of other errors in the span of $\{E_{ig}\}$.
The recovery channel $\C R$ will map states in $\C T$ and $\C J$ (also $\C S$) back to states in $\C T$.
In other words, we see that the modification to the standard condition~(\ref{eq:qec})
is to replace square error operators by rectangular ones,
and increase the number of them by a factor of $d_{\C J}^2$.

Unitary logical operators $U$ on subsystem codes not only commute with $P$,
but also factorize as \be UP=(U_{\C T}\otimes \I_{\C J})P, \label{eq:subgate}\ee
for a nontrivial gate $U_{\C T}$.
The Eastin-Knill theorem holds for subsystem codes,
while note that in the proof the projector is $P$ instead of $P_{\C T}$,
and the actions on the gauge part are ignored.
For subsystem quasi codes,
there still could be a nontrivial logical gate $e^{i\xi PBP}$ on the code subspace $\C T$,
in general.

\newpage


\bibliography{ext}{}

\begin{thebibliography}{10}

\bibitem{Shor96}
Peter~W Shor.
\newblock Fault-tolerant quantum computation.
\newblock In {\em Proceedings of 37th Conference on Foundations of Computer
  Science}, pages 56--65. IEEE, 1996.

\bibitem{AB97}
D.~Aharonov and M.~Ben-Or.
\newblock Fault-tolerant quantum computation with constant error.
\newblock In {\em Proceedings of the 29th annual ACM symposium on Theory of
  computing}, pages 176--188. ACM, 1997.

\bibitem{KLZ98}
E.~Knill, R.~Laflamme, and W.~H. Zurek.
\newblock Resilient quantum computation: error models and thresholds.
\newblock {\em Proc. Royal Soc. A: Mathematical, Physical and Engineering
  Sciences}, 454(1969):365--384, 1998.

\bibitem{Pre98}
John Preskill.
\newblock Reliable quantum computers.
\newblock {\em Proc. Roy. Soc. A: Mathematical, Physical and Engineering
  Sciences}, 454(1969):385--410, Jan 1998.

\bibitem{KLZ96}
E.~Knill, R.~Laflamme, and W.~Zurek.
\newblock Threshold accuracy for quantum computation, 1996.
\newblock arXiv:quant-ph/9610011.

\bibitem{EK09}
B.~Eastin and E.~Knill.
\newblock Restrictions on transversal encoded quantum gate sets.
\newblock {\em Phys. Rev. Lett.}, 102:110502, Mar 2009.

\bibitem{ZCC11}
Bei Zeng, Andrew Cross, and Isaac~L Chuang.
\newblock Transversality versus universality for additive quantum codes.
\newblock {\em IEEE Trans. Inf.}, 57(9):6272--6284, 2011.

\bibitem{CCC+08}
Xie Chen, Hyeyoun Chung, Andrew~W. Cross, Bei Zeng, and Isaac~L. Chuang.
\newblock Subsystem stabilizer codes cannot have a universal set of transversal
  gates for even one encoded qudit.
\newblock {\em Phys. Rev. A}, 78:012353, Jul 2008.

\bibitem{Got98}
Daniel Gottesman.
\newblock Theory of fault-tolerant quantum computation.
\newblock {\em Phys. Rev. A}, 57:127--137, Jan 1998.

\bibitem{Kit03}
A~Yu Kitaev.
\newblock Fault-tolerant quantum computation by anyons.
\newblock {\em Ann. Phys.}, 303(1):2--30, 2003.

\bibitem{BK13}
Sergey Bravyi and Robert K\"onig.
\newblock Classification of topologically protected gates for local stabilizer
  codes.
\newblock {\em Phys. Rev. Lett.}, 110:170503, Apr 2013.

\bibitem{PY15}
Fernando Pastawski and Beni Yoshida.
\newblock Fault-tolerant logical gates in quantum error-correcting codes.
\newblock {\em Phys. Rev. A}, 91:012305, Jan 2015.

\bibitem{BBK+16}
Michael~E Beverland, Oliver Buerschaper, Robert Koenig, Fernando Pastawski,
  John Preskill, and Sumit Sijher.
\newblock Protected gates for topological quantum field theories.
\newblock {\em J. Math. Phys.}, 57(2):022201, 2016.

\bibitem{BK05}
Sergey Bravyi and Alexei Kitaev.
\newblock Universal quantum computation with ideal clifford gates and noisy
  ancillas.
\newblock {\em Phys. Rev. A}, 71:022316, Feb 2005.

\bibitem{PR13}
Adam Paetznick and Ben~W Reichardt.
\newblock Universal fault-tolerant quantum computation with only transversal
  gates and error correction.
\newblock {\em Phys. Rev. Lett.}, 111(9):090505, 2013.

\bibitem{KL96}
E.~Knill and R.~Laflamme.
\newblock Concatenated quantum codes, 1996.
\newblock arXiv:quant-ph/9608012.

\bibitem{JL14}
Tomas Jochym-O’Connor and Raymond Laflamme.
\newblock Using concatenated quantum codes for universal fault-tolerant quantum
  gates.
\newblock {\em Phys. Rev. Lett.}, 112(1):010505, 2014.

\bibitem{NSS+08}
Chetan Nayak, Steven~H Simon, Ady Stern, Michael Freedman, and Sankar~Das
  Sarma.
\newblock Non-abelian anyons and topological quantum computation.
\newblock {\em Rev. Mod. Phys.}, 80(3):1083, 2008.

\bibitem{KKR10}
Robert Koenig, Greg Kuperberg, and Ben~W Reichardt.
\newblock Quantum computation with turaev--viro codes.
\newblock {\em Ann. Phys.}, 325(12):2707--2749, 2010.

\bibitem{BD12}
NE~Bonesteel and DP~DiVincenzo.
\newblock Quantum circuits for measuring levin-wen operators.
\newblock {\em Phys. Rev. B}, 86(16):165113, 2012.

\bibitem{ZHB17}
Guanyu Zhu, Mohammad Hafezi, and Maissam Barkeshli.
\newblock Quantum origami: Transversal gates for quantum computation and
  measurement of topological order, 2017.
\newblock arXiv:quant-ph/1711.05752.

\bibitem{LNC+97}
Debbie~W Leung, Michael~A Nielsen, Isaac~L Chuang, and Yoshihisa Yamamoto.
\newblock Approximate quantum error correction can lead to better codes.
\newblock {\em Phys. Rev. A}, 56(4):2567, 1997.

\bibitem{CGS05}
Claude Cr{\'e}peau, Daniel Gottesman, and Adam Smith.
\newblock Approximate quantum error-correcting codes and secret sharing
  schemes.
\newblock In {\em Annual International Conference on the Theory and
  Applications of Cryptographic Techniques}, pages 285--301. Springer, 2005.

\bibitem{RW05}
M.~Reimpell and R.~F. Werner.
\newblock Iterative optimization of quantum error correcting codes.
\newblock {\em Phys. Rev. Lett.}, 94:080501, Mar 2005.

\bibitem{BO10}
C\'edric B\'eny and Ognyan Oreshkov.
\newblock General conditions for approximate quantum error correction and
  near-optimal recovery channels.
\newblock {\em Phys. Rev. Lett.}, 104:120501, Mar 2010.

\bibitem{BO11}
C\'edric B\'eny and Ognyan Oreshkov.
\newblock Approximate simulation of quantum channels.
\newblock {\em Phys. Rev. A}, 84:022333, Aug 2011.

\bibitem{Ben11}
C\'edric B\'eny.
\newblock Perturbative quantum error correction.
\newblock {\em Phys. Rev. Lett.}, 107:080501, Aug 2011.

\bibitem{NM10}
Hui~Khoon Ng and Prabha Mandayam.
\newblock Simple approach to approximate quantum error correction based on the
  transpose channel.
\newblock {\em Phys. Rev. A}, 81:062342, Jun 2010.

\bibitem{MN12}
Prabha Mandayam and Hui~Khoon Ng.
\newblock Towards a unified framework for approximate quantum error correction.
\newblock {\em Phys. Rev. A}, 86:012335, Jul 2012.

\bibitem{Pre00}
John Preskill.
\newblock {Quantum clock synchronization and quantum error correction}, 2018.
\newblock arXiv:quant-ph/0010098.

\bibitem{HNP+17}
Patrick Hayden, Sepehr Nezami, Sandu Popescu, and Grant Salton.
\newblock {Error Correction of Quantum Reference Frame Information}, 2018.
\newblock arXiv:quant-ph/1709.04471.

\bibitem{FCS+18}
Fernando G. S.~L. Brand\~ao, Elizabeth Crosson, M.~Burak \ifmmode
  \mbox{\c{S}}\else \c{S}\fi{}ahino\ifmmode~\breve{g}\else \u{g}\fi{}lu, and
  John Bowen.
\newblock Quantum error correcting codes in eigenstates of
  translation-invariant spin chains.
\newblock {\em Phys. Rev. Lett.}, 123:110502, Sep 2019.

\bibitem{FNA+19}
Philippe Faist, Sepehr Nezami, Victor~V. Albert, Grant Salton, Fernando
  Pastawski, Patrick Hayden, and John Preskill.
\newblock {Continuous symmetries and approximate quantum error correction},
  2019.
\newblock arXiv:quant-ph/1902.07714.

\bibitem{WA19}
Mischa~P. Woods and \'{A}lvaro~M. Alhambra.
\newblock {Continuous groups of transversal gates for quantum error correcting
  codes from finite clock reference frames}, 2019.
\newblock arXiv:quant-ph/1902.07725.

\bibitem{Note1}
We shall remark that there may not be a sharp distinction between tunable (or
  controllable) parameters and untunable ones. Such a boundary can depend on
  technology, for instance.

\bibitem{DN06}
Christopher~M Dawson and Michael~A Nielsen.
\newblock The {S}olovay-{K}itaev algorithm.
\newblock {\em Quantum Inf. Comput.}, 6(1):81--95, 2006.

\bibitem{AKLT87}
Ian Affleck, Tom Kennedy, Elliott~H. Lieb, and Hal Tasaki.
\newblock Rigorous results on valence-bond ground states in antiferromagnets.
\newblock {\em Phys. Rev. Lett.}, 59:799--802, Aug 1987.

\bibitem{GR07}
Martin Greiter and Stephan Rachel.
\newblock Valence bond solids for $\mathrm{SU}(n)$ spin chains: Exact models,
  spinon confinement, and the haldane gap.
\newblock {\em Phys. Rev. B}, 75:184441, May 2007.

\bibitem{KHK08}
Hosho Katsura, Takaaki Hirano, and Vladimir~E Korepin.
\newblock Entanglement in an su(n) valence-bond-solid state.
\newblock {\em J. Phys. A: Math. Theor.}, 41(13):135304, 2008.

\bibitem{MUM+14}
Takahiro Morimoto, Hiroshi Ueda, Tsutomu Momoi, and Akira Furusaki.
\newblock $z_3$ symmetry-protected topological phases in the su(3) aklt model.
\newblock {\em Phys. Rev. B}, 90:235111, Dec 2014.

\bibitem{CGW11}
Xie Chen, Zheng-Cheng Gu, and Xiao-Gang Wen.
\newblock Classification of gapped symmetric phases in one-dimensional spin
  systems.
\newblock {\em Phys. Rev. B}, 83:035107, Jan 2011.

\bibitem{SPC11}
Norbert Schuch, David P\'erez-Garc\'{i}a, and Ignacio Cirac.
\newblock Classifying quantum phases using matrix product states and projected
  entangled pair states.
\newblock {\em Phys. Rev. B}, 84:165139, Oct 2011.

\bibitem{CGL+13}
Xie Chen, Zheng-Cheng Gu, Zheng-Xin Liu, and Xiao-Gang Wen.
\newblock Symmetry protected topological orders and the group cohomology of
  their symmetry group.
\newblock {\em Phys. Rev. B}, 87:155114, Apr 2013.

\bibitem{DQ13a}
Kasper Duivenvoorden and Thomas Quella.
\newblock Topological phases of spin chains.
\newblock {\em Phys. Rev. B}, 87:125145, Mar 2013.

\bibitem{DQ13b}
Kasper Duivenvoorden and Thomas Quella.
\newblock From symmetry-protected topological order to landau order.
\newblock {\em Phys. Rev. B}, 88:125115, Sep 2013.

\bibitem{RB01}
Robert Raussendorf and Hans~J. Briegel.
\newblock A one-way quantum computer.
\newblock {\em Phys. Rev. Lett.}, 86:5188--5191, May 2001.

\bibitem{GE07}
D.~Gross and J.~Eisert.
\newblock Novel schemes for measurement-based quantum computation.
\newblock {\em Phys. Rev. Lett.}, 98:220503, May 2007.

\bibitem{WSR17}
Dong-Sheng Wang, David~T. Stephen, and Robert Raussendorf.
\newblock Qudit quantum computation on matrix product states with global
  symmetry.
\newblock {\em Phys. Rev. A}, 95:032312, Mar 2017.

\bibitem{SWP+17}
David~T. Stephen, Dong-Sheng Wang, Abhishodh Prakash, Tzu-Chieh Wei, and Robert
  Raussendorf.
\newblock Computational power of symmetry-protected topological phases.
\newblock {\em Phys. Rev. Lett.}, 119:010504, Jul 2017.

\bibitem{KL97}
E.~Knill and R.~Laflamme.
\newblock Theory of quantum error-correcting codes.
\newblock {\em Phys. Rev. A}, 55:900--911, Feb 1997.

\bibitem{KSV02}
A.~Kitaev, A.~H. Shen, and M.~N. Vyalyi.
\newblock {\em Classical and Quantum Computation}, volume~47 of {\em Graduate
  Studies in Mathematics}.
\newblock American Mathematical Society, Providence, 2002.

\bibitem{Wat18}
John Watrous.
\newblock {\em The theory of quantum information}.
\newblock Cambridge University Press, 2018.

\bibitem{GLN05}
Alexei Gilchrist, Nathan~K. Langford, and Michael~A. Nielsen.
\newblock Distance measures to compare real and ideal quantum processes.
\newblock {\em Phys. Rev. A}, 71:062310, Jun 2005.

\bibitem{Sch96}
Benjamin Schumacher.
\newblock Sending entanglement through noisy quantum channels.
\newblock {\em Phys. Rev. A}, 54:2614--2628, Oct 1996.

\bibitem{puncture}
Note that there is a subtlety regarding the efficient computability of
  numbers~\cite{BV97}, so that the whole group needs to be replaced by a
  `punctured' version, here we find we can ignore this without affecting our
  conclusion.

\bibitem{cell}
We assume $\I_\eta$ is unitary. If $\I_\eta$ is not unitary, we may use
  distance measures on channels, such as the infidelity and diamond norm, and
  the results still hold.

\bibitem{PVW+07}
D~Perez-Garcia, F~Verstraete, MM~Wolf, and JI~Cirac.
\newblock Matrix product state representations.
\newblock {\em Quantum Inf. Comput.}, 7(5):401--430, 2007.

\bibitem{Sch11}
Ulrich Schollw{\"o}ck.
\newblock The density-matrix renormalization group in the age of matrix product
  states.
\newblock {\em Ann. Phys.}, 326(1):96--192, 2011.

\bibitem{WAR18}
Dong-Sheng Wang, Ian Affleck, and Robert Raussendorf.
\newblock Topological qubits from valence bond solids.
\newblock {\em Phys. Rev. Lett.}, 120:200503, May 2018.

\bibitem{KLP05}
David Kribs, Raymond Laflamme, and David Poulin.
\newblock Unified and generalized approach to quantum error correction.
\newblock {\em Phys. Rev. Lett.}, 94:180501, May 2005.

\bibitem{Pou05}
David Poulin.
\newblock Stabilizer formalism for operator quantum error correction.
\newblock {\em Phys. Rev. Lett.}, 95:230504, Dec 2005.

\bibitem{NP07}
Michael~A. Nielsen and David Poulin.
\newblock Algebraic and information-theoretic conditions for operator quantum
  error correction.
\newblock {\em Phys. Rev. A}, 75:064304, Jun 2007.

\bibitem{BV97}
Ethan Bernstein and Umesh Vazirani.
\newblock Quantum complexity theory.
\newblock {\em SIAM Journal on Computing}, 26(5):1411--1473, 1997.

\end{thebibliography}
\bibliographystyle{unsrt}


\end{document}